\documentclass[a4paper,10pt,twoside]{cpc-hepnp}

\usepackage{multicol}
\usepackage{graphicx}
\usepackage{booktabs}
\usepackage{amssymb,bm,mathrsfs,bbm,amscd}
\usepackage[tbtags]{amsmath}
\usepackage{lastpage}
\usepackage{CJK}

\usepackage[noblocks]{authblk}

\usepackage{multirow}

\begin{document}
\begin{CJK*}{GBK}{song}

\fancyhead[r]{\large Submitted to Chinese Physics C }

%%\footnotetext[0]{Received 14 January 2013}

\title{The CDEX-1 1 kg Point-Contact Germanium Detector for Low Mass Dark Matter Searches}
%%\thanks{Supported by National Natural Science Foundation of China
%%    (10935005, 10945002, 11275107, 11175099) and National Basic Research program
%%    of China(973 Program) (2010CB833006)}}

\author[1]{KANG Ke-Jun\protect\footnotemark[3]}
\author[1]{YUE Qian\protect\footnotemark[3]\protect\footnotemark[2]}
\author[1]{WU Yu-Cheng\protect\footnotemark[3]}
\author[1]{CHENG Jian-Ping\protect\footnotemark[3]}
\author[1]{LI Yuan-Jing\protect\footnotemark[3]}
\author[3]{BAI Yang\protect\footnotemark[3]}
\author[5]{BI Yong\protect\footnotemark[3]}
\author[4]{CHANG Jian-Ping\protect\footnotemark[3]}
\author[1]{CHEN Nan\protect\footnotemark[3]}
\author[1]{CHEN Ning\protect\footnotemark[3]}
\author[1]{CHEN Qing-Hao\protect\footnotemark[3]}
\author[6]{CHEN Yun-Hua\protect\footnotemark[3]}
\author[7]{CHUANG You-Chun}
\author[1]{DENG Zhi\protect\footnotemark[3]}
\author[1]{DU Qiang\protect\footnotemark[3]}
\author[1]{GONG Hui\protect\footnotemark[3]}
\author[1]{HAO Xi-Qing\protect\footnotemark[3]}
\author[1]{HE Qing-Ju\protect\footnotemark[3]}
\author[3]{HU Xin-Hui\protect\footnotemark[3]}
\author[2]{HUANG Han-Xiong\protect\footnotemark[3]}
\author[7]{HUANG Teng-Rui}
\author[1]{JIANG Hao\protect\footnotemark[3]}
\author[7]{LI Hau-Bin}
\author[1]{LI Jian-Min\protect\footnotemark[3]}
\author[1]{LI Jin\protect\footnotemark[3]}
\author[4]{LI Jun\protect\footnotemark[3]}
\author[2]{LI Xia\protect\footnotemark[3]}
\author[3]{LI Xin-Ying\protect\footnotemark[3]}
\author[3]{LI Xue-Qian\protect\footnotemark[3]}
\author[1]{LI Yu-Lan\protect\footnotemark[3]}
\author[7]{LIAO Heng-Ye}
\author[7]{LIN Fong-Kay}
\author[7]{LIN Shin-Ted}
\author[5]{LIU Shu-Kui\protect\footnotemark[3]}
\author[1]{LV Lan-Chun\protect\footnotemark[3]}
\author[1]{MA Hao\protect\footnotemark[3]}
\author[4]{MAO Shao-Ji\protect\footnotemark[3]}
\author[1]{QIN Jian-Qiang\protect\footnotemark[3]}
\author[2]{REN Jie\protect\footnotemark[3]}
\author[1]{REN Jing\protect\footnotemark[3]}
\author[2]{RUAN Xi-Chao\protect\footnotemark[3]}
\author[6]{SHEN Man-Bin\protect\footnotemark[3]}
\author[7,8]{Lakhwinder SINGH}
\author[7,8]{Manoj Kumar SINGH}
\author[7,8]{Arun Kumar SOMA}
\author[1]{SU Jian\protect\footnotemark[3]}
\author[5]{TANG Chang-Jian\protect\footnotemark[3]}
\author[7]{TSENG Chao-Hsiung}
\author[6]{WANG Ji-Min\protect\footnotemark[3]}
\author[5]{WANG Li\protect\footnotemark[3]}
\author[1]{WANG Qing\protect\footnotemark[3]}
\author[7]{WONG Tsz-King Henry}
\author[6]{WU Shi-Yong\protect\footnotemark[3]}
\author[3]{WU Wei\protect\footnotemark[3]}
\author[4]{WU Yu-Cheng\protect\footnotemark[3]}
\author[5]{XING Hao-Yang\protect\footnotemark[3]}
\author[3]{XU Yin\protect\footnotemark[3]}
\author[1]{XUE Tao\protect\footnotemark[3]}
\author[1]{YANG Li-Tao\protect\footnotemark[3]}
\author[7]{YANG Song-Wei}
\author[1]{YI Nan\protect\footnotemark[3]}
\author[3]{YU Chun-Xu\protect\footnotemark[3]}
\author[1]{YU Hao\protect\footnotemark[3]}
\author[5]{YU Xun-Zhen\protect\footnotemark[3]}
\author[6]{ZENG Xiong-Hui\protect\footnotemark[3]}
\author[1]{ZENG Zhi\protect\footnotemark[3]}
\author[4]{ZHANG Lan\protect\footnotemark[3]}
\author[6]{ZHANG Yun-Hua\protect\footnotemark[3]}
\author[3]{ZHAO Ming-Gang\protect\footnotemark[3]}
\author[1]{ZHAO Wei\protect\footnotemark[3]}
\author[3]{ZHONG Su-Ning\protect\footnotemark[3]}
\author[2]{ZHOU Zu-Ying\protect\footnotemark[3]}
\author[5]{ZHU Jing-Jun\protect\footnotemark[3]}
\author[4]{ZHU Wei-Bin\protect\footnotemark[3]}
\author[1]{ZHU Xue-Zhou\protect\footnotemark[3]}
\author[6]{ZHU Zhong-Hua\protect\footnotemark[3]}

\footnotetext[2]{Corresponding author. E-mail: yueq@mail.tsinghua.edu.cn}
\footnotetext[3]{The member of CDEX Collaboration}

\affil[1]{Department of Engineering Physics, Tsinghua University, Beijing, 100084}
\affil[2]{Institute of Nuclear Physics, China Institute of Atomic Energy, Beijing, 102413}
\affil[3]{School of Physics, Nankai University, Tianjin, 300071}
\affil[4]{NUCTECH Company, Beijing, 100084}
\affil[5]{School of Physical Science and Technology, Sichuan University, Chengdu, 610065}
\affil[6]{YaLong River Hydropower Development Company, Chengdu, 610051}
\affil[7]{Institute of Physics, Academia Sinica, Taipei, 11529}
\affil[8]{Department of Physics, Banaras Hindu University, Varanasi, 221005}

\maketitle

%% Abstract
\begin{abstract}
  The CDEX Collaboration has been established for direct detection of light dark
  matter particles, using ultra-low energy threshold p-type point-contact
  germanium detectors, in China JinPing underground Laboratory (CJPL). The first
  1 kg point-contact germanium detector with a sub-keV energy threshold has been
  tested in a passive shielding system located in CJPL. The outputs from both
  the point-contact p$^+$ electrode and the outside n$^+$ electrode make it
  possible to scan the lower energy range of less than 1 keV and at the same
  time to detect the higher energy range up to 3 MeV. The outputs from both
  p$^+$ and n$^+$ electrode may also provide a more powerful method for signal
  discrimination for dark matter experiment. Some key parameters, including
  energy resolution, dead time, decay times of internal X-rays, and system
  stability, have been tested and measured. The results show that the 1 kg
  point-contact germanium detector, together with its shielding system and
  electronics, can run smoothly with good performances. This detector system
  will be deployed for dark matter search experiments.
\end{abstract}

%% Keywords

%% PACS: refer to http://www.aip.org/pacs/pacs.html/

%%\footnotetext[0]{\hspace*{-3mm}\raisebox{0.3ex}{$\scriptstyle\copyright$}2013
%%  Chinese Physical Society and the Institute of High Energy Physics of the
%%  Chinese Academy of Sciences and the Institute of Modern Physics of the Chinese
%%  Academy of Sciences and IOP Publishing Ltd}%

\begin{multicols}{2}

\section{Introduction}
\label{sec:intro}

Light dark matter particles with masses of less than 10 GeV have been come a new
target for direct detection experiments. In order to search for dark matter
Weakly Interacting Massive Particles (WIMP) in the low mass region, it is
necessary to develop a detector system with an ultra-low energy threshold, as
well as keeping its background level ultra-low. High Purity Germanium (HPGe) has
been chosen as the target and detector for dark matter searches due to its very
low radioactivity, very good energy resolution, ultra-low energy threshold and
modular structure, which makes it easy to scale up to larger and larger masses
of detector array while keeping almost the same performances as that of a small
mass detector module. The China Dark matter Experiment (CDEX) Collaboration was
established in 2009 to start a new program for searching for light dark matter,
using ultra-low energy threshold germanium array detector systems. The physical
goals and technical feasibility of the CDEX experiment were explored some time
ago before the collaboration itself \cite{YueQ2004}. The first physics results
for a dark matter search with an ultra-low energy threshold HPGe detector in a
surface laboratory were published by the TEXONO collaboration \cite{LinST2009},
and are also partially based on such endeavors.

Other experiments, such as CoGeNT \cite{Aaleth2011}, XENON \cite{Aprile2012},
CDMS \cite{Ahmed2011}, CRESST \cite{Angloher2012}, DAMA \cite{Bernabei2010} and
so on, scan the low mass region for dark matter with different targets based on
different technologies. The most stringent exclusive curve so far has been given
by the XENON experiment in 2012. The WIMP-nucleon spin-independent cross-section
is about $10^{-45}$ $\mbox{cm}^2$ at a WIMP mass of 50 GeV, but the sensitivity
is still not good in the low mass region of less than 10 GeV
\cite{Angloher2012}, even though the results from XENON have excluded the regions
claimed by CoGeNT, DAMA and CRESST. The results from CoGeNT, DAMA and CRESST are
also inconsistent with each other. All of these new results show us that the
searching for WIMP in the low mass region has become topic of keen debate in recent years.

The CDEX collaboration will directly detect light dark matter particle WIMPs
with masses of about 10 GeV using a tonne-scale germanium detector array
composed of many 1 kg-scale PCGe (point-contact germanium) detectors. The energy
threshold level and other performances of the tonne-scale detector should be
almost the same, therefore, as that of a 1 kg-scale detector. A 1kg-scale PCGe
detector which can achieve an ultra-low energy threshold of less than 500 eV
makes it powerful enough to scan the low mass region of dark matter. As a first
step, the CDEX collaboration has studied a 1 kg PPCGe detector (CDEX-1).

\section{CJPL}
\label{sec:cjpl}

China JinPing underground Laboratory (CJPL) is located in the central part
of a 17.5 km-long traffic tunnel which was built for the construction of hydropower
plants on both sides of JinPing Mountain in Sichuan province, southwest China.
The rock overburden in the central part of the traffic tunnel is about 2400 m
(6720 m water equivalent depth). The construction of CJPL started in 2009 and the
laboratory has been formally running since Dec. 2010. The current volume of CJPL is
about 4000 m$^3$ \cite{Kang2010}. The cosmic-ray flux has been measured by two triple-coincident
scintillation counter telescopes and the Muon flux measured to be about 60 muons
y$^{-1}$m$^{-2}$ \cite{WuYC2013}. This low flux is highly beneficial for dark
matter searches and other rare event experiments in situ. Both the deep rock
overburden to shield from cosmic-ray and the ambient rock with very low
radioactivity make CJPL the best underground laboratory in the world for
ultra-low background experiments such as dark matter, double beta decay and so
on. It is also planned to further enlarge the space available at CJPL to host
more experiments in the future.

\section{The CDEX-1 detector system}
\label{sec:cdex1}

The CDEX-1 detector system has been set up in a polyethylene room, which has 1
meter-thick wall inside the CJPL. The whole system
consists of three parts: a 1 kg point-contact germanium detector; electronics
and read out system; and a shielding system. This paper will describe the
structure and performance of CDEX-1 in situ at CJPL.

\subsection{1 kg-scale PPCGe detector}
\label{sec:1kg-det}

The point-contact technology of a HPGe detector was developed several decades ago
based on the general coaxial germanium detector technology \cite{Luke1989}. In
order to achieve an ultra-low energy threshold, the area of the germanium
detector electrode should be as small as possible. We know that beside
electronics noise the noise of a detector depends mainly on the capacitance of
the detector. The capacitance of a detector is mainly related to the size of the
electrode on the germanium detector. The electrode size of the contact point on
a germanium detector can reach mm-scale and the corresponding capacitance can be
$\sim 1$ pF. This point-contact technology provides the possibility to decrease
the energy threshold of a germanium detector down to 500 eV or even lower.

Collaborating with Canberra Company, the CDEX collaboration has developed a
p-type point-contact germanium (PPCGe) detector with a mass of 1 kg, which is
the largest in the world so far. The structural materials of the 1 kg PPCGe
detector, with ultra-low radioactivity background, and its upgraded low noise
pre-amp electronics give the detector a low radiation background and low noise
qualities. The structure of the CDEX-1 PPCGe is shown in
Fig.\ref{fig:det-struct}. The 1 kg PPCGe crystal is encapsulated within 1.5
mm-thick Oxygen Free High Conductivity (OFHC) copper endcap. To suppress
external low energy gamma rays (and X-rays), the endcap is windowless. The
distance between the germanium crystal and the endcap is about 4mm. The crystal
cylinder has an n$^+$ type contact on the outer surface and a tiny p$^+$ type
contact as the central electrode. The small diameter of the central electrode
with the order of 1 mm reduces the capacitance of the detector to the order of 1
pF and greatly improves the intrinsic noise characteristics.

%% Fig.1. The structure of CDEX-1 1 kg PPCGe detector.
\begin{center}
  \includegraphics[width=7cm]{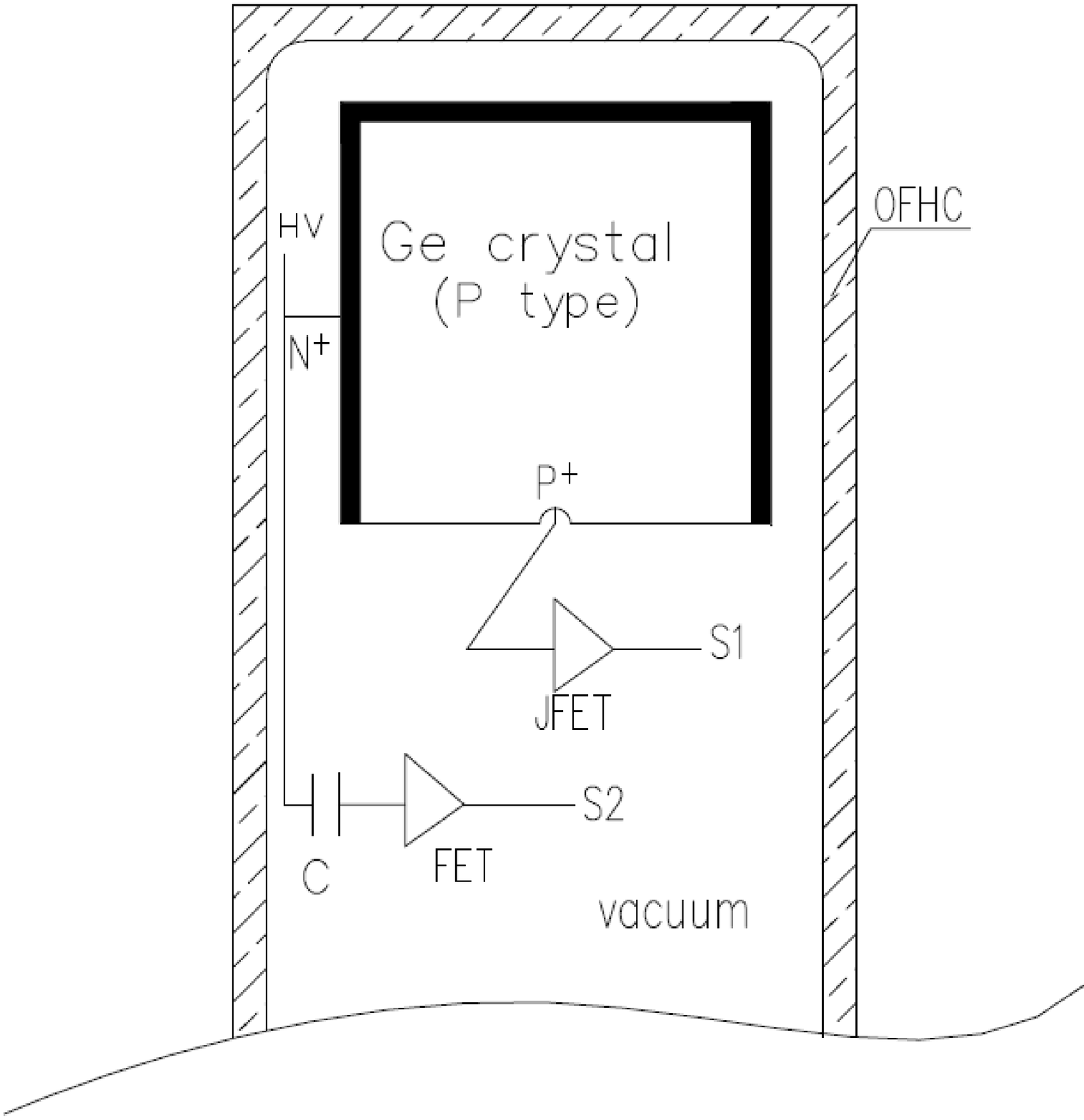}
  \figcaption{\label{fig:det-struct} The structure of CDEX-1 1 kg PPCGe
    detector.}
\end{center}

\subsection{The CDEX-1 electronics and read out system}
\label{sec:cdex1-ro}

The pre-amplifier outputs of the CDEX-1 1 kg PPCGe detector include the S1
signal from the point-contact p$^+$ electrode and S2 signal from the n$^+$
electrode which also served as the HV electrode. The p$^+$ point contact signal
is readout by a pulsed optical feedback preamplifier with an ultra-low noise
JFET nearby, and the signal from the n$^+$ type electrode is also readout by a
resistive feedback preamplifier. Each pre-amplifier has 4 outputs: three
identical \verb|OUT_E| for energy measurement and one \verb|OUT_T| for timing
measurement which was not used for this experiment. The multiple outputs provide
more choices to connect more main amplifiers with different shaping times and
gains. All outputs should be well connected to the high impedance inputs of
downstream modules. The detector is recommended to be operated under +3500 V
high voltage.

The electronics and data acquisition system of the CDEX-1 1 kg PPCGe is
illustrated simply in Fig.\ref{fig:electronics}. All the NIM/VME modules and
crates remain commercial products from Canberra \cite{Canberra} and CAEN
\cite{CAEN} companies. In order to distinguish different pulse shapes, the
signals from one \verb|OUT_E| of each preamplifier are amplified by a
fast timing amplifier (Canberra 2111) and then fed into the flash
analog-to-digital converter (FADC, CAEN V1724, 100MHz sampling frequency) for
fast pulse digital processing. The other preamplifier outputs are directly
connected into a conventional spectroscopy amplifier (Canberra 2026) and then
fed into the FADC for digitization. The signals from n$^+$ electrode are also
fed into the FADC for digitization. One signal from the p$^+$ point contact
electrode is discriminated after the spectroscopy amplifier and served as one of
the trigger for the detector system. The random trigger signals at rate of 0.05
Hz from a pulse generator are used to measure the dead time of the electronics
and read out system. All the data is transferred to a PC through a duplex
optical fiber. The total trigger rate of the CDEX-1 1 kg PPCGe detector system
is kept less than 10 Hz for long term data taking.

%% Big picture: Fig.2 The schematic electronic diagram of CDEX-1 1 kg PPCGe detector
\end{multicols}
%%\ruleup

\begin{center}
  \includegraphics[width=15cm]{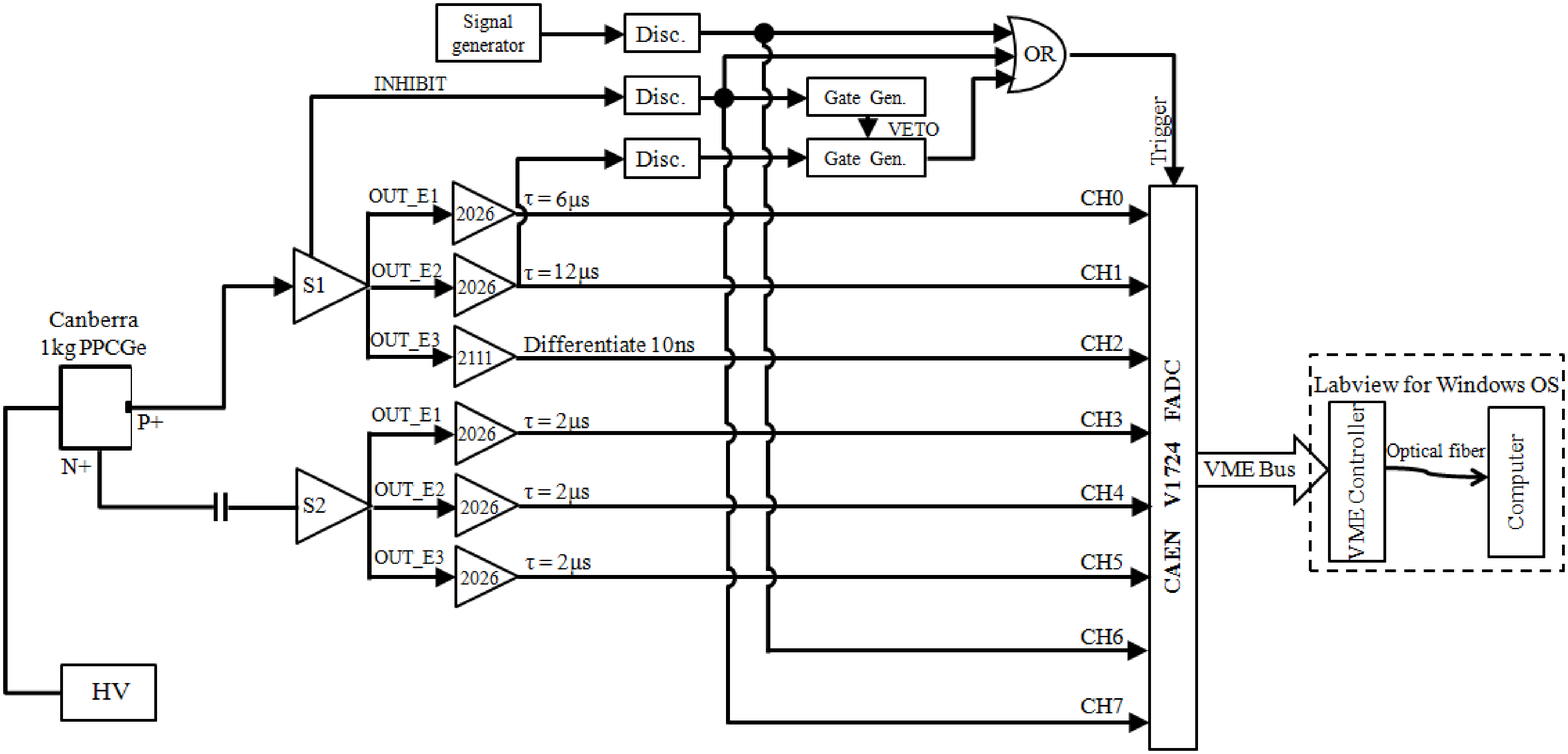}
  \figcaption{\label{fig:electronics} The schematic electronic diagram of CDEX-1
    1 kg PPCGe detector}
\end{center}
%%\ruledown

\begin{multicols}{2}

\subsection{The shielding system}
\label{sec:cdex1-shielding}

The CDEX-1 1 kg PPCGe detector was installed into CJPL in order to avoid
background from cosmic-rays. A passive shielding system has been setup for
shielding from gamma ray or neutron backgrounds from ambient rock and
materials. The structure of the shielding system is shown in
Fig.\ref{fig:shielding} and the materials from outside to inside are: 20 cm
thick lead to shield from external gamma radiation from rock and other
materials; and 20 cm thick boron-loaded polyethylene for neutron deceleration
and thermal neutron absorption. The whole shielding system is located inside a 1
m thick layer of polyethylene for neutron shielding, which is not shown in
Fig.\ref{fig:shielding}. The CDEX-1 1 kg PPCGe detector was housed in the
shielding system along with LN2 Dewar. A 20 cm thick layer of OFHC copper
surrounds the cryostat of the PPCGe detector to further decrease the residual
gamma background from outside. The internal space between the 20 cm OFHC copper
shielding and the cryostat flushed with pure nitrogen gas to eliminate
radioactive radon gas.

%% Fig.3 The shielding system of CDEX-1
\begin{center}
  \includegraphics[width=7cm]{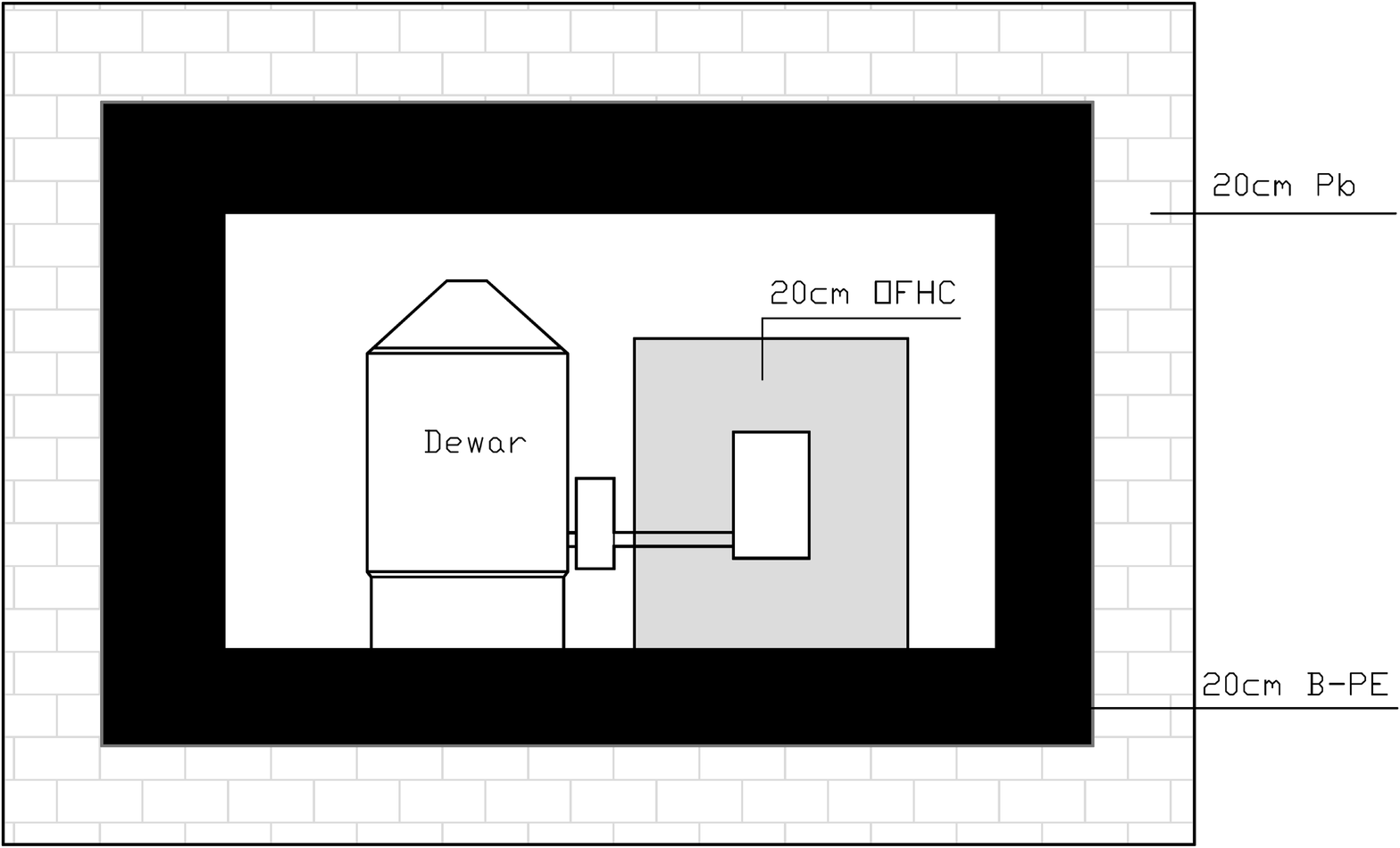}
  \figcaption{\label{fig:shielding} The shielding system of CDEX-1}
\end{center}

\section{The CDEX-1 1 kg PPCGe detector Performances}
\label{sec:performances}

The CDEX-1 1 kg PPCGe detector has been installed and thoroughly tested to
achieve its optimal performances. Due to the 1.5 mm OFHC copper window, the 1 kg
PPCGe detector could not be calibrated with external low-energy gamma or X-rays,
as they cannot pass through the OFHC copper window.  The studies of the detector
performances have to be done using its own internal characteristic X-ray lines
in the low energy range.

The measured spectrum includes the 10.37 keV K-shell X-ray from $^{71}$Ge and $^{68}$Ge
atoms, the 8.98 keV X-ray of $^{65}$Zn atom and even 1.29 keV L-shell X-ray from $^{71}$Ge
and $^{68}$Ge atoms. These internal characteristic X-ray lines can be used to
calibrate the detector, monitor the stability of the detector system and study
the energy resolution at different energy ranges. Many of the characteristic
X-ray lines at the energy range of less than 12 keV are summarized in Table
\ref{tab:X-rays}.

%% Tab.1 The energies and lifetimes of K-shell and L-shell X-rays for different atoms.
\begin{center}
  \tabcaption{\label{tab:X-rays} The energies and lifetimes of K-shell and L
   -shell X-rays for different atoms.}\footnotesize
  \begin{tabular*}{80mm}{c@{\extracolsep{\fill}}ccc}
    \toprule {\bf Atomic}      &  {\bf E (K-shell)}  &  {\bf E (L-shell)}  &  {\bf Lifetime}  \\
    species     &    (keV)  &    (keV)  &     (day)  \\ \hline
    $^{73}$As        &    11.10  &    1.414  &    80.30  \\ \hline
    $^{71}$Ge        &    10.37  &    1.298  &    11.43  \\ \hline
    $^{68}$Ge        &    10.37  &    1.298  &    270.8  \\ \hline
    $^{68}$Ga        &     9.66  &    1.194  &    0.047  \\ \hline
    $^{65}$Zn        &     8.98  &    1.096  &    244.3  \\ \hline
    $^{56}$Ni        &     7.71  &    0.926  &    6.077 \\ \hline
    $^{56}$Co        &           &           &    271.9  \\
    $^{57}$Co        &     7.11  &    0.846  &   77.28  \\
    $^{58}$Co        &           &           &    70.87  \\ \hline
    $^{55}$Fe        &     6.54  &    0.769  &    997.1  \\ \hline
    $^{54}$Mn        &     5.99  &    0.695  &    312.3  \\ \hline
    $^{51}$Cr        &     5.46  &    0.628  &    27.70  \\ \hline
    $^{49}$V         &     4.97  &    0.564  &      330  \\ \hline
  \end{tabular*}
\end{center}

%%\hphantom{00}

\subsection{Energy calibration}
\label{sec:eng-cali}

As illustrated in Fig.\ref{fig:electronics}, different gains and shaping times
are chosen to process the pre-amplifier signals. The channels from S1 have been
set to only cover the low energy range below 12 keV and they can only be
calibrated by the intrinsic characteristic X-ray lines of the natural long-life
cosmogenic radioactive nuclei. Meanwhile, the channels
from S2 are used to trace the backgrounds in relatively higher energy
regions. So the channels from S2 are calibrated by some radiation source
samples, e.g. Europium in our cases. Many peaks are fitted to do the energy
calibration and energy linearity study. The 10.37 keV peak and its fit result
are shown in Fig.\ref{fig:x-res} as a sample. The calibration results of
different channels are displayed in Fig.\ref{fig:eng-cali}, showing one channel
from S1 and three channels from S2. The calibration information can be seen in
Tab.\ref{tab:X-rays}. At the same time, the zero energy point can be defined
with random trigger events.

%% Fig.4 10.37 keV K-shell X-ray peak of 68,71Ge and the energy resolution fitting result.
\begin{center}
  \includegraphics[width=7cm]{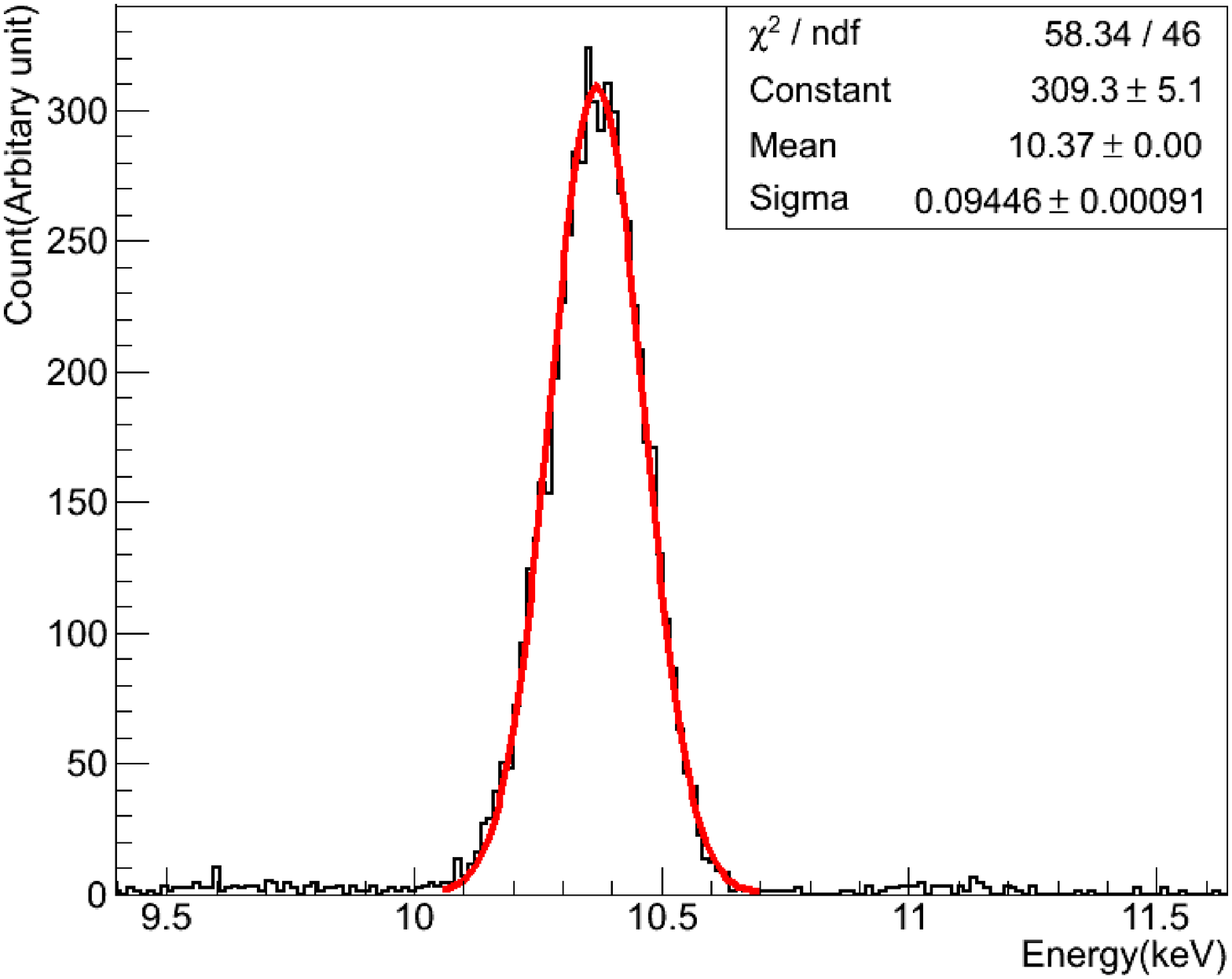}
  \figcaption{\label{fig:x-res} The 10.37 keV K-shell X-ray peak of $^{68,71}$Ge and
    the energy resolution fitting result.}
\end{center}

%% Fig.5. Energy calibrations of CDEX-1 1 kg PPCGe detector including both S1 \verb|OUT_E2| channel (a) and three channels from S2 with different gains (b) (c) (d).
\begin{center}
  \includegraphics[width=7cm]{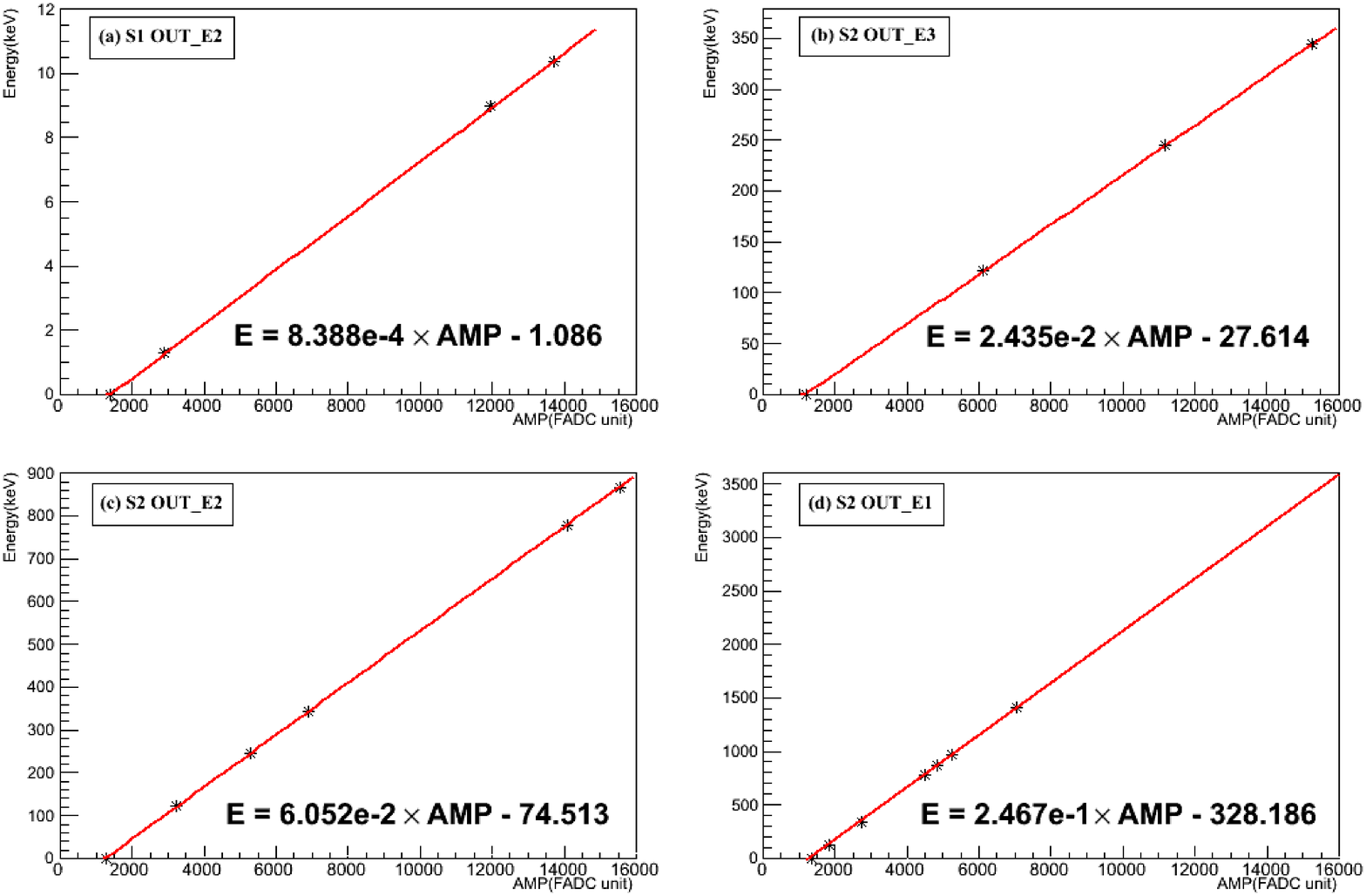}
  \figcaption{\label{fig:eng-cali} Energy calibrations of the CDEX-1 PPCGe
    detector including both S1 channel (a) and three channels from S2
    with different gains (b) (c) (d).}
\end{center}

%% Fig.6. Spectra associated with S1 channel (a) and the channels from S2 (b) (c) (d) with different gains.
\begin{center}
  \includegraphics[width=7cm]{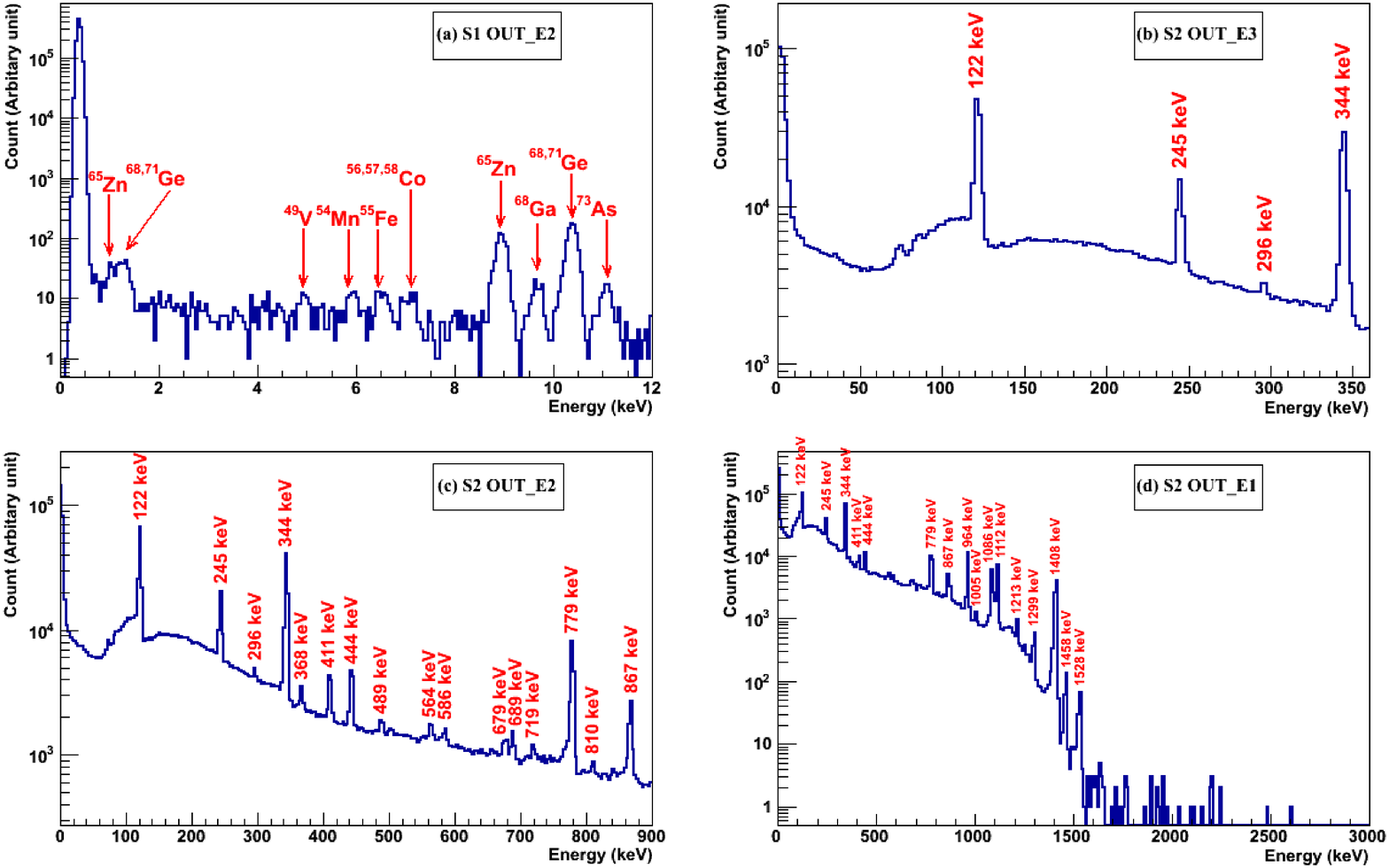}
  \figcaption{\label{fig:spec-gains} Spectra associated with S1  channel (a) and
    the channels from S2 (b) (c) (d) with different gains.}
\end{center}

%% Lang Table: Tab.2  Selection of X-ray and gamma lines for calibration
\end{multicols}

\begin{center}
  \tabcaption{\label{tab:x-cali} Selection of X-ray and gamma lines for
    calibration} \footnotesize
  \begin{tabular*}{170mm}{c|c|c|c|c|c|c|c|c}
    \toprule
    \multicolumn{3}{c|}{S1 OUT\_{}E2} & \multicolumn{2}{c|}{S2 OUT\_{}E3} &
    \multicolumn{2}{c|}{S2 OUT\_{}E2} & \multicolumn{2}{c|}{S2 OUT\_{}E1} \\ \hline
    \multicolumn{3}{c|}{Background} & \multicolumn{6}{c|}{Source $^{152}$Eu} \\ \hline
    Isotope     &  E (keV)  &  FWHM (keV)  &  E (keV)  &  FWHM (keV)  &  E (keV)  &  FWHM (keV)  &  E (keV)  &  FWHM (keV)    \\ \hline
      $^{68,71}$Ge (X)  & 1.299 &0.251     &  121.8   & 4.44  &   121.8  &       5.82  &   121.8  &  7.47       \\ \hline
      $^{65}$Zn (X)    & 8.979  &0.212     &  244.7   & 4.81  &   244.7  &       7.02  &   344.3  &  6.39       \\ \hline
      $^{68,71}$Ge (X)  & 10.37 &0.222     &  344.3   & 4.19  &   344.3  &       5.34  &   778.9  &  8.04       \\ \hline
                  &          &             &          &       &   778.9  &       6.39  &   867.4  &  9.60       \\ \hline
                  &          &             &          &       &   867.8  &       7.52  &   964.1  &  7.65       \\ \hline
                  &          &             &          &       &          &             &    1408  &  7.00       \\ \hline
  \end{tabular*}
\end{center}

\begin{multicols}{2}

\subsection{Energy resolution}
\label{sec:eng-res}

The measured energy spectra for calibration are shown in
Fig.\ref{fig:spec-gains}. Various characteristic X-ray and gamma lines can be
clearly seen. The resolutions of different energy peaks are calculated and given
in Tab.\ref{tab:x-cali}. One can see from the spectra that
there are many other characteristic X-ray peaks in the S1 \verb|OUT_E2| spectrum and
that the energy threshold can be brought down to less than 1 keV without any
electronic noise suppression. The X-ray peak from $^{65}$Zn L-shell can also be
identified. So, we can see from the background spectrum of S1 \verb|OUT_E2| that the
detector has an ultra-low energy threshold and good energy resolution.

\subsection{Decay of the low-energy characteristic X-rays}
\label{sec:X-decay}

%% Fig.7. Decays of characteristic X-rays associated with $^71$Ge (T1/2 = 11.4 d).
\begin{center}
  \includegraphics[width=7cm]{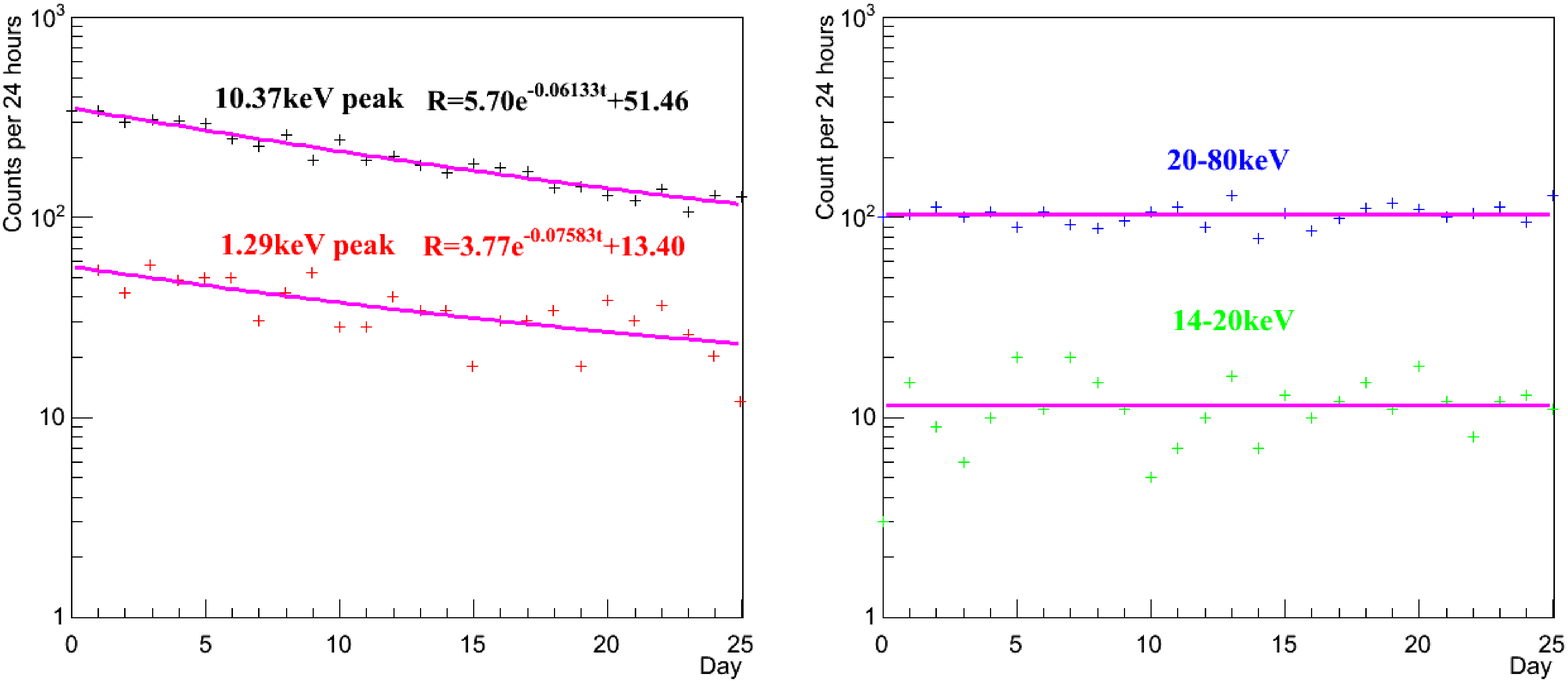}
  \figcaption{\label{fig:x-decay} Decays of characteristic X-rays associated
    with $^{71}$Ge ($T_{1/2} = 11.4$ d).}
\end{center}

The characteristic X-rays observed in the low-energy spectrum from the
point-contact electrode of the 1 kg PPCGe detector originate from the cosmogenic
activation of the germanium crystal. After a long time of exposure to cosmic
rays at ground level the intensities of these X-rays will achieve a balanced
status. After the germanium detector was moved into CJPL, the balanced status
was broken and the number of radioactive nuclei decreased due to the lower
production rate, which is related to the much lower Muon flux inside CJPL. One
can then measure the decays of some short life-time radioactive
nuclei. Fig.\ref{fig:x-decay} shows the decays of the 10.37 keV K-shell EC X-ray
(KX-ray) peak and the 1.29 keV L-shell EC X-ray peak (LX-ray) from $^{68}$Ge and
$^{71}$Ge. Due to the relatively long half-life of $^{68}$Ge, the decay is mainly induced
by the $^{71}$Ge isotope. The rate data is fitted with exponential decay plus a
constant background. The decay times of the 10.37 keV peak ($11.3 \pm 0.9$ d)
and the 1.29 keV peak ($11.8 \pm 3.0$ d) are coincident with the half-life of
$^{71}$Ge ($T_{1/2} = 11.4$ d). At the same time, there are no clear decays for
events in the 14-20 keV and 20-80 keV energy ranges, which are mainly due to
long-lived radioactive isotopes and other backgrounds.

\subsection{Dead time}
\label{sec:deadtime}

To calculate the real event rate, it is necessary to know the dead time of the
data acquisition (DAQ) system of CDEX-1 1 kg PPCGe detector. One can see that in
Fig.\ref{fig:electronics}, there is a signal generator which contributes about
0.05 Hz to the total trigger rate of the CDEX-1 DAQ system. This periodic
trigger can be considered as independent from the physical triggers from the 1
kg PPCGe detector and so be served as random triggers. The dead time of CDEX-1
DAQ system can then be calculated as the ratio of the unrecorded random trigger
number and the generated random trigger number. Based on the data already
collected the dead time of the CDEX-1 DAQ system is less than 1\%.

%% Fig.8.  Stability check of the CDEX-1 PPCGe detector.
\begin{center}
  \includegraphics[width=7cm]{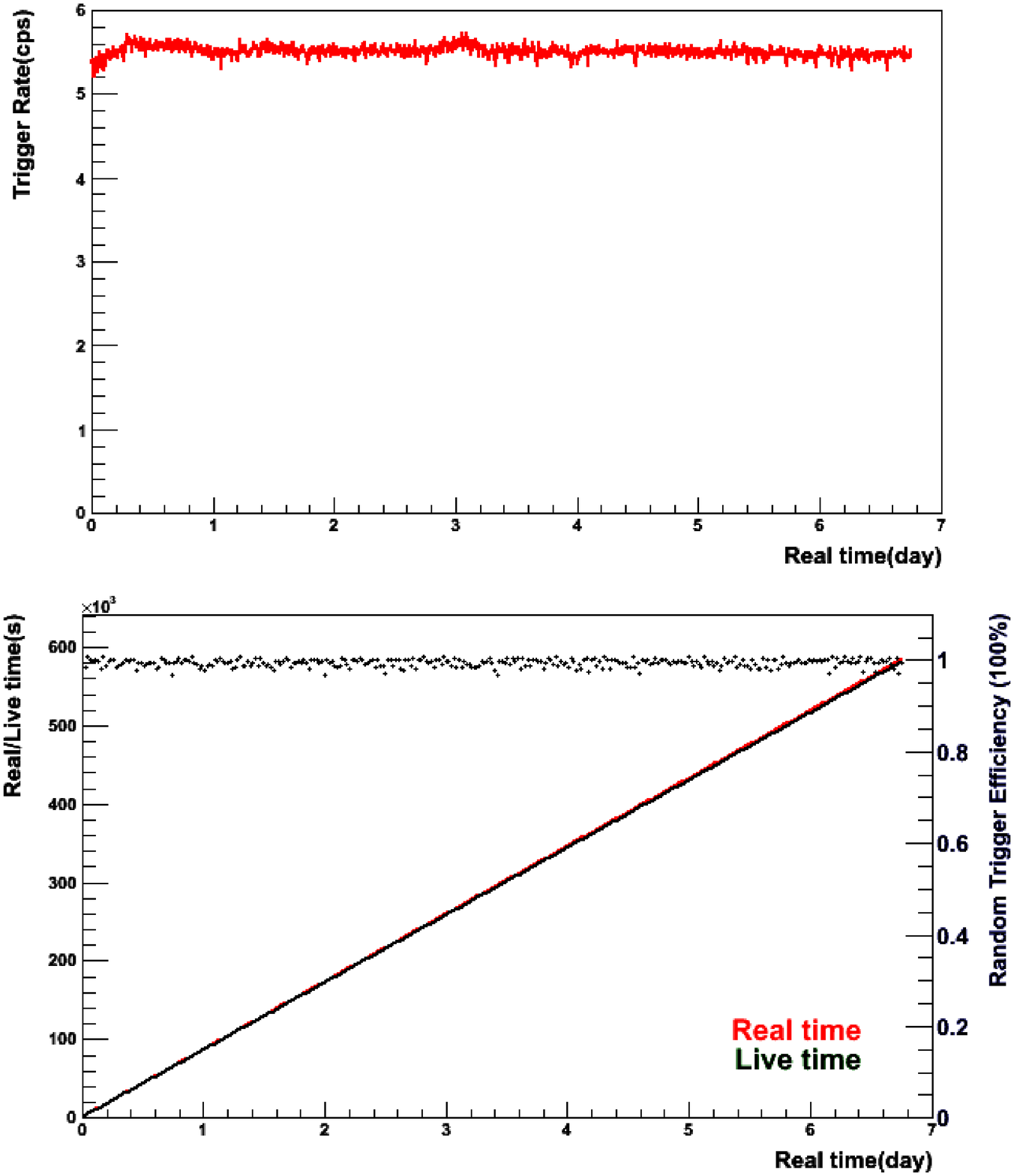}
  \figcaption{ \label{fig:stablility} Stability check of the CDEX-1 PPCGe
    detector.}
\end{center}

\subsection{Stability of the 1 kg PPCGe detector}
\label{sec:stability}

To verify the validity of the data, several preliminary offline analysis were
carried out, including trigger rate, random trigger efficiency and the ratio of
real time to live time. In Fig.\ref{fig:stablility}, one can see that the
trigger rate and random trigger efficiency of the CDEX-1 DAQ are both relatively
stable, showing that we can expect the whole experiment system to run smoothly
and stably.

\section{Summary}
\label{sec:sum}

The CDEX collaboration has been established to search for dark matter particles
with a tonne-scale mass germanium detector array system. As the first stage
experiment, the CDEX collaboration has set up a point-contact germanium detector
with a mass of 1 kg scale (CDEX-1) and studied the performance of the whole
system in CJPL. The results show that the CDEX-1 detector system can run
smoothly with good energy resolution and an ultra-low energy threshold. For the
next step this detector system will be used to directly search for dark matter
particle WIMPs and we look forward to physics results being available soon.

\vspace{10mm}

This work is Supported by National Natural Science Foundation of China
(10935005, 10945002, 11275107, 11175099) and National Basic Research program
of China(973 Program) (2010CB833006).

%% End main body
\end{multicols}

%% 分隔线

\vspace{-1mm}
\centerline{\rule{80mm}{0.1pt}}
\vspace{2mm}

%% 参考文献

\begin{multicols}{2}

\end{multicols}

\clearpage

\end{CJK*}

\begin{thebibliography}{90}
\vspace{3mm}

%[1]
\bibitem[]{YueQ2004} YUE Qian, CHENG Jian-Ping, LI Yuan-Jing {\it et al.}
  HEP\&NP, 2004, {\em 28(8)}: 877-880; HE Dao, LI Yu-Lan, YUE Qian {\it et al.}
  HEP\&NP, 2004, {\em 30(6)}: 548-533; LI Xin, YUE Qian, LI Yuan-Jing {\it et al.}
  HEP\&NP, 2007, {\em 31(6)}: 564-569
%[2]
\bibitem{LinST2009} LIN Shin-Ted, LI Hau-Bin, LI Xin {\it et al.} Phys. Rev. D,
  2009, {\em 79}, 061101(R)
%[3]
\bibitem{Aaleth2011} Aaleth C. E. {\it et al.} (CoGeNT Collaboration),
  Phys. Rev. Lett., 2011, {\em 107}, 141301
%[4]
\bibitem{Aprile2012} Aprile E. {\it et al.} (XENON 100 Collaboration),
  Phys. Rev. Lett., 2012, {\em 109}, 181301
%[5]
\bibitem[]{Ahmed2011} Ahmed Z {\it et al.} (CDMS Collaboration),
  Phys. Rev. Lett., 2011, {\em 106}, 131302
%[6]
\bibitem[]{Angloher2012} Angloher G {\it et al.} (CRESST Collaboration),
  Eur. Phys. J. C, 2012, {\em 72}, 1971
%[7]
\bibitem{Bernabei2010} Bernabei R {\it et al.} (DAMA Collaboration),
  Eur. Phys. J. C, 2010, {\em 67}, 39; Belli P {\it et al.} Phys. Rev. D, 2011, {\em
    84}, 055014
%[8]
\bibitem[]{Kang2010} KANG Ke-Jun, CHENG Jian-Ping, CHEN Yun-Hua {\it et al.}
  Journal of Physics: Conference Series, 2010, {\em 203}: 012028
%[9]
\bibitem{WuYC2013} WU Yu-Cheng, HAO Xi-Qing, YUE Qian {\it et al.} Measurement
  of Cosmic Ray Flux in China JinPing underground Laboratory. Chin. Phys. C, to
  be published
%[10]
\bibitem{Luke1989} Luke P {\it et al.} IEEE Trans. Nucl. Sci. 36, {\em 926}
  (1989)
%[11]
\bibitem[]{Canberra} http://www.canberra.com
%[12]
\bibitem{CAEN} http://www.caen.it

\end{thebibliography}
\end{document}